\documentclass[twocolumn,amsmath,amssymb,preprintnumbers]{revtex4}

\usepackage{graphicx}
\usepackage{epsfig}
\begin{document}

\title{Automatic anomaly detection in high energy collider data}

\author{Simon de Visscher}
\email[E-mail: ]{simon.de.visscher@cern.ch}
\affiliation{Physik-Institut, Universit\"{a}t Z\"{u}rich, Winterthurerstrasse 190, CH-8057, Z\"{u}rich, Switzerland}

\author{Michel Herquet}
\email[E-mail: ]{mherquet@nikhef.nl}
\affiliation{Nikhef Theory Group, Science Park 105, 1098 XG Amsterdam, The Netherlands}

\preprint{NIKHEF-2010-012}

\date{\today}

\begin{abstract}
We address the problem of automatic anomaly detection in high energy collider data. Our approach is based on the random generation of analytic expressions for kinematical variables, which can then be evolved following a genetic programming procedure to enhance their discriminating power. We apply this approach to three concrete scenarios to demonstrate its possible usefulness, both as a detailed check of reference Monte-Carlo simulations and as a model independent tool for the detection of New Physics signatures.
 \end{abstract}
\maketitle

\section{Introduction}
The analysis of data produced by existing and forthcoming high energy collider experiments is notoriously challenging, both from theoretical and experimental perspectives. The situation worsens with hadron collisions at high luminosity due their intrinsic complexity and the huge amount of events to be considered. Existing efforts to face those challenges and isolate signal signatures above predominant backgrounds generally belong to either of two categories.  On the one hand, top-down approaches take advantage of the prior knowledge of specific theoretical framework (e.g., the Standard Model, or a particular supersymmetric scenario) to predict and simulate interesting signatures in order define advanced and efficient isolation strategies. Those strategies can then be applied to real data and the results readily interpreted in terms of (model dependent) constraints on a given parameter space. On the other hand, bottom-up approaches first focus on data through simple physical observables (e.g., invariant masses, sum of transverse momenta) in order to identify possible deviations from background expectations. 

Two important remarks can be made at this stage. First in both cases our ability to precisely simulate backgrounds and thus, indirectly, to carefully validate Monte-Carlo predictions from various sources, is a crucial ingredient for discovery. Second, the nature of some of the most relevant signals could very well be such that both types of approach initially fail to identify them. This is particularly true if relevant signatures do not present enough similarities with those studied in the context of existing top-down analysis, and, at the same time, are sufficiently complex not to lead to statistically significant variations in the limited set of physical observables considered in bottom-up approaches.

The present work aims at addressing those questions by introducing a new method to automatically detect and characterize anomalies (i.e., statistically significant deviations from a given reference) both in Monte-Carlo samples and real data. This method makes use of the \textit{genetic programming} technique introduced by Koza~\cite{Koza92} (see \cite{poli08:fieldguide} for a global overview and \cite{teodorescu-2008} for examples of applications in High Energy Physics \footnote{In High Energy Physics, genetic programming was, to our knowledge, only used for signal dependent cut optimization (e.g, see \cite{Cranmer:physics0402030} and \cite{Link1,Link2}). Genetic algorithms, a similar yet different technique, have been also been used for large-scale parameter optimization and fitting problems (e.g., see \cite{Abdullin:2006nu}, \cite{Allanach:2004my} and \cite{DelDebbio:2007ee,Ball:2008by}).}), and used recently to automatically ``derive'' fundamental conservation laws from chaotic pendulum data \cite{MichaelSchmidt04032009}. Contrary to top-down approaches, no prior assumption is made on the nature of the anomalies to be isolated and, contrary to bottom-up approaches, the technique is not limited to a small subset of physical quantities but can take advantage of the discriminating property of more complex observables.

This letter is organized into three parts. First, we define the anomaly detection method based on the genetic programming technique. Second, we apply this method to three simple yet relevant analysis scenarios at the LHC to demonstrate its usefulness both as a test of Monte-Carlo simulations and as a model independent tool for early stage New Physics discovery. Finally, we comment on the obtained results and discuss prospects to generalize our approach to a wider class of final state configurations.

\section{The method}
\begin{figure}[]
\includegraphics[width=4cm]{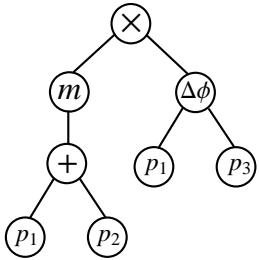}
\caption{Syntax tree for the example formula $\varphi=m(p_1+p_2)\Delta\phi(p_1,p_3)$, where $p_i$ are ordered particle 4-momenta and $m$ and $\Delta\phi$ represent respectively the invariant mass and the difference in azimuthal angles functions. The function and terminal sets are in this case $\{+,\times,m,\Delta\phi\}$ and $\{p_1,p_2,p_3\}$. The depth, i.e., the length of the longest branch, is 4.}
\label{fig:tree}
\end{figure}

We start by defining a set (also called \textit{population}) of \textit{formulas} $\varphi_i$ randomly built on a set of \textit{terminals} (typically, variables such as the four-momenta of reconstructed objects in the detectors) and a set of \textit{functions} (e.g., addition, invariant mass). Each of those formula is represented by a \textit{syntax tree}, as shown in Figure \ref{fig:tree}. 
The initial method for the random generation of the formula set is arbitrary, but a popular choice is a mixture of the so-called ``full'' and ``grow'' algorithms leading to trees with branches of constant or varying lengths. Each function and terminal is \textit{strongly typed} such that the resulting formulas are built by combining quantities with physical units in a meaningful way.

After the initialization procedure, the \textit{fitness} of the generated formulas, i.e., their ability to highlight discrepancies between a considered set of data and a given sample of events considered as a reference, is evaluated. Assuming that both the data and the reference sample are normalized to a certain number of events and, for a given formula $\varphi$, are distributed over a certain range in $n$ bins of equal lengths denoted $d[l]$ and $r[l]$, we propose to use
\begin{equation}
\label{eq:chi2}
\chi^2_\varphi=\sum_{l=1}^n \left(\frac{d[l]-r[l]}{\sqrt{d[l]}}\right)^2
\end{equation}
as a naive estimator of the formulas $\varphi$ fitness when the statistical uncertainty over the reference $r[l]$ is negligible. If the formulas $\varphi_i$ are random and independent, and the number of considered events sufficiently large, this estimator initially follows, by definition, a $\chi^2$ distribution with $n$ degrees of freedom. In practice, in order to deal with bins in which the observed number of signal events is very low  (i.e., below a certain threshold $\sigma_{min}^2$), the denominator of expression \eqref{eq:chi2} can be replaced by $max(\sqrt{d[l]},\sigma_{min})$ to avoid artificially large contributions.

The next step is to make the initial population evolve. A new population of identical size is generated by creating new formulas according to three distinct reproduction mechanisms. First, given two parent formula, the \textit{subtree crossover} randomly and independently selects a crossover node in each parent tree and creates an offspring by replacing the subtree rooted at the crossover point in the first parent with the subtree rooted at the crossover point in the second parent. In order to avoid the creation of offsprings with arbitrarily long depths, the initial node selection can be adjusted such that the offspring depth always remains smaller than a fixed maximal value. Second, the \textit{mutation} which corresponds to a crossover between a parent formula and another formula, external to the initial population and generated randomly. Finally, the \textit{copy} mechanism exactly reproduces a parent formula.

Each mechanism has a certain probability to happen $p_{cross}$, $p_{mut}$ and $p_{copy}$, with, typically, $p_{cross}\gg p_{mut} > p_{copy}$. To ensure a constant selection pressure, parents with an higher fitness are selected more frequently to take part in the reproduction. The presence of three distinct reproduction mechanisms satisfies three crucial requirements. Formulas with the best discrepancy power should share their ``genetic inheritance'' more often, they should have the opportunity to exactly duplicate (to avoid the disappearance of the best candidates) and a certain amount of randomness should be present to decrease the sensitivity on initial conditions (i.e., an initial population with a limited size).

The fitness evaluation and evolution steps are repeated alternatively during a given fixed number of iterations. The best formula is identified for each iteration, and the best candidate for \textit{all} iterations is reported. Alternatively, the whole last population, or a subset of it, can also be reported as worth considering alternatives. The fitness of the best individual(s) $\chi_{\varphi}^{2(best)}$ allows us to estimate of the actual significance of the observed discrepancy(ies). First, the $p$-value $p_{best}$ associated to $\chi_{\varphi}^{2(best)}$ is calculated assuming a normal $\chi_{\varphi}^2$ distribution with $n$ degrees of freedom. This value cannot be directly interpreted as a statistical deviation since the total population, i.e., all the formulas generated during $N_{iter}$ iterations, is not randomly distributed and biased by construction towards high $\chi^2$ values. To circumvent this difficulty, we calculate the $p$-value $p_{ref}$ associated with the value $\chi_{\varphi}^{2(ref)}$ obtained by applying multiple times the same algorithm on a signal-free reference sample containing the same number of events as the data, and by averaging the results. Finally, we propose to use the ratio $p_{best}$/$p_{ref}$, which corresponds to the probability of obtaining the value $\chi_{\varphi_i}^2\geq\chi_{\varphi}^{2(best)}$ in a subsample of formulas with  $\chi_{\varphi_i}^2\geq\chi_{\varphi}^{2(ref)}$ as an estimate of the discrepancy significance.

\section{Applications}
To illustrate the proposed technique, we apply it to three different simple analysis scenarios. Applicability of the method to a wider class of objects and to more realistic data are discussed in the next section. The results have been produced using a prototype genetic programing library \cite{GenHEP} running on a single desktop machine. The algorithm parameters are kept constant to emphasize process independence and allow or an easy comparison of the results. The population size is 50. Formulae have an initial depth of 4 which can increase up to 5. The number of iteration is fixed to 10 and the reproduction mechanism probabilities defined in the previous section are set to $p_{cross}=0.75$, $p_{mut}=0.20$ and $p_{copy}=0.05$. The terminals are the considered particle four-momenta and and the function set is
\begin{equation}
f = \{+,\times,m,\Delta \eta, \Delta \phi, \Delta R,\cos\theta\}
\end{equation}
where $m$ is the invariant mass, $\Delta\eta$ and $\Delta\phi$ the differences in pseudo-rapidity and azimutal angle, $\Delta R\equiv\sqrt{\Delta\eta^2+\Delta\phi^2}$ and $\cos\theta$ is the cosine of the angle between the particle direction and the beam axis. The fitness estimator \eqref{eq:chi2} is computed for 10 bin distributions. To estimate the statistical error for the output, it has been applied 100 times on different samples and the (gaussian distributed) results have been summarized in terms of mean and standard deviation.

 \subsection{Spin correlation in top quark decay}
Our first example aims at demonstrating how the method can be used to quantitatively compare two Monte-Carlo simulations supposedly describing the same process(es). We confront two samples containing top quark pairs decaying into a full leptonic final state, namely $pp\rightarrow t\overline{t}\rightarrow \mu^+ \nu_\mu b \mu^-\overline{\nu}_\mu \overline{b}$. The parton-level of the first sample is simulated with MadGraph/MadEvent (MG/ME) \cite{Alwall:2007st} using the exact $2\rightarrow 6$ matrix element. The second sample is obtained using Pythia~\cite{Sjostrand:2006za} only, with the exact $2\rightarrow 2$ matrix element for the top quark pair production, but generating the six particle final state through a flat phase space decay. Obviously, the two samples would gave similar results for simple quantities such as transverse momenta or invariant masses. However, they would differ significantly when considering spin dependent observables such as particular angular distributions.

Applying the method described above, we obtain a fitness estimator of $47.7\pm 9.8$ when comparing the exact $2\rightarrow 6$ matrix element sample with a flat phase space decay sample, to be compared to a value of $12.3\pm 2.2$ when comparing two different samples implementing the same flat phase space decay. This translates into a probability larger than 99\% that the two samples are based on physically incompatible descriptions of the same process. Note that the most discriminating individuals all contain explicit references to angular variable distributions such as $\Delta\phi$, pointing out to the exact nature of the discrepancy.
 
 \subsection{$W'$ boson and SUSY decay chain}
In the following examples, we consider two cases. The first is the $s$-channel production of  a heavy (1 TeV) $W'$ gauge boson, decaying as $W'\rightarrow WZ\rightarrow \mu^\pm\mu^-\mu^+\nu$ to give a particularly clean three muons final state. The second process considered yields an identical final state, i.e., three leptons, but it originates from a longer, supersymmetric decay chain $pp\rightarrow (\tilde{\chi}_1^+\rightarrow \nu_\mu\tilde{\mu}^+_L\rightarrow \nu_\mu \mu^+ \tilde{\chi}_1^0)(\tilde{\chi}_2^0\rightarrow \mu^-\tilde{\mu}^+_R\rightarrow \mu^- \mu^+ \tilde{\chi}_1^0)$, where we assumed $m_{\chi_1^+}=m_{\chi_2^0}=500$~GeV, $m_{\tilde{\mu}_L^+}=202$~GeV and  $m_{\tilde{\mu}_R^-}=144$~GeV. 

 The method is applied to the comparison of two samples, each containing a mixture of one of these BSM signals with the associated irreducible SM background, with a variable $S/(S+B)$ ratio, and a sample containing only the SM background $pp\rightarrow WZ\rightarrow \rightarrow \mu^\pm\mu^-\mu^+\nu $. The event production consists of to the parton-level output of MG/ME. The relation
 \begin{figure}[]
\includegraphics[width=0.49\textwidth]{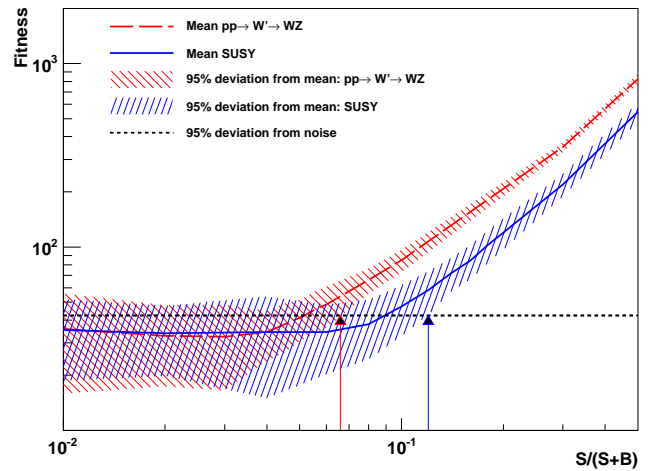}
\caption{Fitness as a function of $S/(S+B)$ for the $W'\rightarrow WZ\rightarrow 3\mu$ (red) and for the $pp\rightarrow (\tilde{\chi}_1^+\rightarrow \nu_\mu\tilde{\mu}^+_L\rightarrow \nu_\mu \mu^+ \tilde{\chi}_1^0)(\tilde{\chi}_2^0\rightarrow \mu^-\tilde{\mu}^+_R\rightarrow \mu^- \mu^+ \tilde{\chi}_1^0)$ (blue) signals. The red (blue) line and the associated band indicate the mean and 95\% deviation for the fitness of the best function respectively. The dotted line indicates the 95\% $p$-value threshold computed using a signal-free sample, as described in the text.}
\label{fig:wp}
\end{figure} 
 between the average fitness of the best individuals after a fixed number of iteration and the $S/(S+B)$ ratio is shown in Figure \ref{fig:wp}. 
 For the $W'$ case the signal remains unvisible below an $S/(S+B)$ ratio of approximatively 7\%. The slightly higher fitness values observed in the 4-6\% region are not large enough to be statistically significant, considering the large variation of the results in the signal free region. Above 7\%, the average fitness is large enough to provide a statistically significant discrimination of the signal over the background. As expected, all the most discriminant distributions contain combinations proportional to the total invariant mass of the three final state leptons.

 In the SUSY case, the presence of a signal can be detected at a 95\% C.L.  for $S/(S+B)>10\%$, which is not very different from the result one would expect from classical cut based studies using similar data. It turns out that the most discriminant distributions are in fact mainly based on the total invariant mass of the three final state leptons, but a large fraction of them also contain information related to their angular distribution. 

\section{Discussion}
We discuss the merits of our proposal in the context of three questions:

\textit{How does it compare with approaches using a fixed set of distributions ?}
As perhaps expected by construction, our approach can help isolating discrepancies with  larger statistical significance than strategies based on a limited fixed set of discriminating distributions~\cite{2009JPhCS.171a2102B}. However, the two types of approach remain complementary with, on one hand, faster calculations and an easier control of systematic uncertainties in the context of real data analysis and, on the other hand, better prospects for detection of anomalies with a complex structure (e.g., involving kinematic information of more than two objects). 

\textit{How does it compare with random searches ?}
A common criticism leveled at genetic programing algorithms is their possibly low efficiency when compared to a simple random search over the space of tree formula. Our tests have confirmed that similar results can be achieved with random searches for simple cases. However, for more complex cases (such as our SUSY example) random searches took significantly longer (5-10 times) CPU time to isolate the best discriminating formula, which often appear longer and unnecessarily more complex. This lets us postulate that particularly complex cases, such as many particle final states, could only be handled by techniques implementing an optimization procedure. Finally, the proposed genetic programming solution allows to optimize a complete set of formula among which different variations can be finally preferred based on simplicity over the most discriminating one, where a random search would provide only a few relevant solutions.

\textit{Can it be generalization to more complex objects with systematic uncertainty propagation ?} The examples we presented only use muon momentum information such that, in first approximation, systematic reconstruction uncertainties can be safely neglected compare to statistical fluctuations. However, when considering more complex objects (jets, missing $E_T$), effects like finite energy and spatial resolutions, reconstruction algorithm limitations, and various other detector effects cannot be overlooked. We would suggest to use transfer functions (such as in Matrix Element based techniques~\cite{madweight} to propagate errors associated, for example, with jet property measurements. The resulting additional error would then be naturally convoluted with the intrinsic statistical error associated with the method. This approach would naturally lead to a serious increase in the calculation complexity, which should however be well-manageable using a dedicated library written using a fast, compiled language. 

\section{Conclusion}
We have proposed a new approach to automatize anomaly detection in high energy collider data. Our method is based on the random generation of analytic expressions for complex kinematical variables, which are then evolved using a genetic programming procedure to enhance their discriminating power. We successfully applied this method in three different test cases and demonstrated its potential usefulness for both Monte Carlo sample validation and BSM anomaly detection. 

Besides those already mentioned, we believe there are several prospects for improvement and generalization of our approach. First, though it is a model independent anomaly detection technique, it might well be helpful at a later stage of a classical ``top down'' analysis to build (instead of guessing) the optimal, most discriminating distribution variable after several acceptance and analysis cuts have been applied. Second, a similar technique could be used to improve the ``blind'' generation of BSM signatures in the context of effective scenarios (e.g., see \cite{ArkaniHamed:2007fw}). Finally, we hope genetic algorithms would become part of pragmatic solutions to determine the best BSM model fitting a restricted set of data in the context of the so-called ``LHC inverse problem''. Work along those lines is already in progress.

\section{Acknowledgements}
We thank Eric Laenen, Claude Amsler, Christophe Delaere and Johan Alwall for useful discussions. MH research has been supported by the Dutch Foundation for Fundamental Research of Matter (FOM) and the Dutch National Organization for Scientific Research (NWO). 

\bibliography{genetic}

\end{document}